\documentclass{aastex}
\usepackage{natbib}
\usepackage{emulateapj5}

\shorttitle{Measuring Stellar Radial Velocities with DFDI}
\shortauthors{Mahadevan et al.}

\begin{document}

\title{Measuring Stellar Radial Velocities with a Dispersed Fixed-Delay Interferometer }

\author{Suvrath Mahadevan\altaffilmark{1}, Julian van Eyken, Jian Ge, Curtis DeWitt, Scott W. Fleming, Roger Cohen, Justin Crepp  \& Andrew Vanden Heuvel }
\affil{Astronomy Department,  University of Florida, 211 Bryant
Space Science Center P.O. Box 112055 Gainesville, FL 32611-2055}
\email{suvrath@astro.ufl.edu}

\altaffiltext{1}{Visiting Astronomer, Kitt Peak National
Observatory, National Optical Astronomy Observatory. KPNO is
operated by AURA, Inc.\ under contract to the National Science
Foundation.}

\begin{abstract}
We demonstrate the ability to measure precise stellar barycentric
radial velocities with the dispersed fixed-delay interferometer
technique using the Exoplanet Tracker (ET), an instrument
primarily designed for precision differential Doppler velocity
measurements using this technique. Our barycentric radial
velocities, derived from observations taken at the KPNO 2.1 meter
telescope, differ from those of Nidever et al. by 0.047 km
s$^{-1}$ (rms) when simultaneous iodine calibration is used, and
by 0.120 km s$^{-1}$ (rms) without simultaneous iodine
calibration. Our results effectively show that a Michelson
interferometer coupled to a spectrograph allows precise
measurements of barycentric radial velocities even at a modest
spectral resolution of R $\sim 5100$. A multi-object version of
the ET instrument capable of observing $\sim$500 stars per night
is being used at the Sloan 2.5 m telescope at Apache Point
Observatory for the Multi-object APO Radial Velocity Exoplanet Large-area
Survey (MARVELS), a wide-field radial velocity survey for
extrasolar planets around TYCHO-2 stars in the magnitude range
$7.6<V<12$. In addition to precise differential velocities, this
survey will also yield precise barycentric radial velocities for
many thousands of stars using the data analysis techniques
reported here. Such a large kinematic survey at high velocity
precision will be useful in identifying the signature of accretion
events in the Milky Way and understanding local stellar kinematics
in addition to discovering exoplanets, brown dwarfs and
spectroscopic binaries.
\end{abstract}


\section{INTRODUCTION}
The discovery of various tidal streams in our galaxy (Ibata et al.
1994; Yanny et al. 2006; Belokurov et al. 2006) challenges the
monolithic collapse formation model (Eggen 1962), but the streams
discovered so far form only a small fraction of the mass of the
galaxy. Large kinematic surveys are necessary to determine if such
merger and accretion events are more common. Radial velocity
surveys using multi-object spectrographs, like RAVE (Steinmetz et
al. 2006), are already producing barycentric radial velocities for
many thousands of stars at a precision of $2-3$ km s$^{-1}$,
enabling studies of galactic dynamics (e.g. Veltz et al. 2008). In
the northern hemisphere the magnitude limited Geneva-Copenhagen
Survey (Nordstrom et al. 2004) has provided radial velocities for
over 14000 bright nearby F and G dwarfs ($V < 8.6$) at a precision
of $100-300$ m s$^{-1}$. Coupled with gravity and chemical
composition derived from the observed spectra, such datasets of
precise barycentric radial velocities are very valuable in
identifying past accretion events and substructure in the galactic
disk (Helmi et al. 2006), finding members of moving groups (Jones
1970) and determining cluster membership.

The ability to determine precise differential radial velocities
has also improved dramatically in the last few years, and planet
search surveys are currently at a precision of $1-3$ m s$^{-1}$
(Rupprecht et al. 2004; Butler et al. 1996). Such planet search
surveys, however, use an arbitrary zero point for the stellar
velocities since only differential velocities are measured. The
true radial velocity of the center-of-mass of a star is a
difficult quantity to measure accurately using spectroscopy since
it involves additional modelling (or assumptions) of effects like
convective blue shift and gravitational redshift (Lindegren \&
Dravins 2003). A quantity that is easier to obtain, in terms of
observables and known physical constants, is the apparent Doppler
shift of the lines in the stellar spectra, as measured in the
solar system barycenter, with the Sun setting the velocity zero
point. This Doppler shift can be converted into a velocity that we
define as the barycentric radial velocity of the star. Very
precise barycentric radial velocities have been measured using
high resolution echelle spectrographs by Udry et al. 1999 (UD99
hereafter) for a number of stable F, G, K dwarfs, as well as by
Nidever et al. 2002 (ND02 hereafter) for a larger sample.

In this paper we demonstrate the ability of a dispersed
fixed-delay interferometer (DFDI) to measure precise barycentric
radial velocities. The Exoplanet Tracker (ET) is a prototype of
this class of instruments, which use a Michelson interferometer
followed by a medium resolution spectrograph working in the first
grating order. Fixed delay interferometers have long been used by
solar astrophysicists to measure solar oscillations (Gorskii \&
Lebedev 1977; Kozhevatov 1983; Harvey et al. 1995). In 1997 D. J.
Erskine proposed adding a spectrograph as a post disperser for
increasing the measurement precision of differential radial
velocities (Erskine \& Ge 2000; Ge, Erskine \& Rushford 2000).
When using this technique sinusoidal interference fringes are
formed in the slit direction at the position of the stellar
absorption lines. A shift in the dispersion direction of the
underlying spectra due to a Doppler shift manifests itself as the
phase shift of the sinusoidal fringes in the slit direction (Ge
2002).

Precise radial velocities can be measured with DFDI even at low
wavelength dispersions since information is also being encoded as
sinusoidal fringes in the slit direction. This results in each
stellar spectrum taking up only a small fraction of the available
number of pixels on the CCD detector used to record the spectrum.
This feature of the DFDI technique is a great advantage since it
has the potential to enable a multi-object instrument (similar to
ET) to simultaneously obtain precise radial velocities for a large
number of stars (Ge 2002; Ge et al. 2002) without the need for an
expensive cross-dispersed high spectral resolution instrument. One
such instrument, the W. M. Keck Exoplanet Tracker, is being used
at the Sloan 2.5 m telescope for the magnitude limited
Multi-object APO Radial Velocity Exoplanet Large-area Survey (MARVELS), a
survey for extrasolar planets around TYCHO-2 stars in the
magnitude range $7.6<V<12$ that is being conducted as part of SDSS
III (Weinberg et al. 2007). This instrument can observe 59 stars
simultaneously, and approximately 500 stars per night. Designed
primarily as a wide-field precision radial velocity survey to
discover short and intermediate period planets, MARVELS will also
provide a dataset of precise barycentric radial velocities for
thousands of late F, G, and K TYCHO-2 stars.

\section{Observations}
The observations presented in this article were conducted with the
single object ET instrument at the Kitt Peak 2.1 meter telescope
in January 2006 as part of an ongoing planet search program. The
ET instrument is linked to the telescope by a 200 $\mu$m optical
fiber, which is fed by the f/8 beam of the 2.1 meter telescope.
The instrument itself is housed in an isolated room in the
basement and consists of a Michelson interferometer in series with
a medium resolution spectrograph operating in the wavelength range
$5000-5640$ \AA , and with a spectral resolution of R $\sim 5100$.
An iodine cell can be inserted into the optical path and serves to
calibrate out instrument drifts. This instrument has successfully
demonstrated its short-term stability by confirming known planets
(e.g. van Eyken et al. 2004) and discovering a short period planet
around the star HD102195 (Ge et al. 2006).

The observing procedure for the ET planet search program is
similar to other precision radial velocity programs that use an
iodine cell to calibrate out the instrument drifts (e.g. Butler et
al. 1996). For each star in the program a single exposure is
acquired without the iodine cell inserted in the beam path. An
exposure of the iodine cell, illuminated by a tungsten lamp, is
also acquired immediately before or after this stellar exposure.
These exposures are referred to as the star and iodine template
respectively. The iodine template is acquired at high signal to
noise (S/N), typically S/N = 150-200 per pixel. The
signal-to-noise on the star template depends on the brightness of
the star and the exposure time and is generally in the range of
S/N = 50-200 per pixel. All subsequent exposures are taken with
the iodine cell inserted in the path of the stellar beam. It is
from these exposures that precision radial velocities are obtained
since the velocity drift of the iodine component tracks the
instrument drift, allowing it be calibrated out. Exposure times
are similar to those used on the stellar template, typically
yielding a S/N $\sim 1/\sqrt{2}$ of that obtained on the stellar
template (since the iodine cell absorbs almost 50\% of the light).

Stable stars and known planet-bearing stars are routinely observed
for calibration as part of the ET planet search program. During
the observing run we were able to obtain regular observations of
the RV stable stars $\eta$ Cas (small linear trend in velocity),
$\tau$ Ceti, and 36 Uma as well as the known planet bearing stars
51 Peg (Mayor \& Queloz 1995), 55 Cnc (McArthur et al. 2004), and
HD4967 (Butler et al. 2003). For six stars, chosen because they
had rms velocity scatter less than 100 m s$^{-1}$ reported in ND02
(HD3674, HD3861, HD12414, HD105405, HD130087, HD134044), we also
acquired stellar templates and iodine templates immediately before
or after the respective stellar templates, although no
star-with-iodine data points were obtained for these six stars.
Typical exposure times varied from 3 minutes for the brightest
star ($\eta$ Cas, $V=3.45$) to 10 minutes for the faintest
(HD130087, $V=7.52$), leading to a photon noise limited radial
velocity error of $3-10$ m s$^{-1}$ for most observations.

The advantage of an isolated, fiber-fed instrument is the
potential for high stability. The total instrument velocity drift
during the observing run (as tracked by the iodine templates) can
be seen in Figure 1. The instrument velocity drift is only 2.5 km
s$^{-1}$ over the entire 12 day observing run and less than 0.5 km
s$^{-1}$ over the first night. This short term stability is
essential for the data analysis technique described in Section
3.1.

\section{Data Analysis Procedure}
The standard ET data analysis pipeline was designed to extract
differential radial velocity for the planet search program. It
does not give the barycentric radial velocity of the star since
the zero point itself is arbitrary. The pipeline processing steps
are described in more detail in van Eyken et al. 2004, and Ge et
al. 2006. Here we discuss the various processing steps briefly,
but we concentrate on the modifications that allow barycentric
radial velocities to be measured.

The data produced by ET were processed using standard IRAF
procedures, as well as software written in Research System Inc.'s
IDL software. The images were corrected for biases, dark current,
and scattered light and then trimmed, illumination corrected, and
low-pass filtered. The visibilities ($V$) and the phases
($\theta$) of the fringes were determined for each channel by
fitting a sine wave to each column of pixels in the slit
direction, with the fringe visibility, $V$,  being defined as
\begin{equation}
V= (I_{max} - I_{min})/(I_{max} + I_{min})
\end{equation}
where $I_{max}$ and $I_{min}$ are the maximum and minimum values
of the fitted sine wave.
 A wavelength calibration was applied using the observed
Thorium-Argon spectra, and the visibility and phase data were
dispersion corrected and re-binned to a log-lambda wavelength
scale. To determine differential velocity shifts in the
star+iodine data (starlight passing through the iodine cell) the
data can, to a good approximation, be considered as a summation of
the complex visibilities (${\bf V}=Ve^{i\theta}$) of the relevant
star ($V_Se^{i\theta_{S_0}}$) and iodine ($V_Ie^{i\theta_{I_0}}$)
templates (van Eyken et al. 2004; Erskine 2003). For small
velocity shifts the complex visibility of the data can be written
as
\begin{equation}
V_D e^{i\theta_D} = V_S
e^{i\theta_{S_0}}e^{i\theta_S-i\theta_{S_0}}
+V_Ie^{i\theta_{I_0}}e^{i\theta_I-i\theta_{I_0}}
\end{equation}
where $V_D, V_S,$ and $V_I$ are the fringe visibilities for a
given wavelength in the star+iodine data, star template and iodine
template respectively, and $\theta_D, \theta_{S_0}$, and
$\theta_{I_0}$ the corresponding measured phases.
 In the presence of real velocity shift of the star and instrument
 drifts the complex visibilities of the star and iodine template best match the
 data with a phase shift of $\theta_S - \theta_{S_0}$ and $\theta_I -
 \theta_{I_0}$ respectively. The iodine is a stable reference and the iodine phase shift tracks the instrument drift. The
difference between star and iodine shifts is the real phase shift
of the star, $\Delta \phi$, corrected for any instrumental drifts
\begin{equation}
\Delta \phi = (\theta_S - \theta_{S_0}) - (\theta_I -
\theta_{I_0})
\end{equation}
This phase shift can be converted to a velocity shift ($\Delta v$)
by a known phase-to-velocity scaling factor that is a function of
the optical delay in the Michelson interferometer ($d$), the
wavelength ($\lambda$) and the speed of light ($c$). These
parameters are connected by the relation
\begin{equation} \Delta v= \frac{c \lambda}{2 \pi  d}
\Delta \phi \label{eqn:delayfringe2}
\end{equation}

In reality the phase shifts of the star and iodine templates are
not the only free parameters needed to fit the observed data. The
Doppler velocity shift of the star induces a wavelength shift of
the entire spectrum and instrument instabilities can lead to pixel
shifts relative to the templates. Finding the correct phase shifts
requires that these wavelength shifts (or pixels shifts) for the
star and iodine templates also be fit simultaneously with the
phase shift parameters.

 To determine barycentric radial velocities we used
the stellar template of a star with known barycentric radial
velocity ($\eta$ Cas) in place of the target star template in
Equation 2. The visibility of the reference star was allowed to
scale to partially compensate for the template mismatch. The
stellar visibility depends on the line depth and width (Ge 2002),
and allowing it to scale is in effect similar to attempting to
match the line depths and widths of the stellar template with that
of the observed star. Similar spectral-morphing techniques have
been used (Johnson et al. 2006) with high-resolution echelle data
from the N2K survey to achieve short term differential velocity
precision of $\sim 5$ m s$^{-1}$ without the need for a stellar
template.

\subsection{Radial Velocity using Templates Alone}

The stability of the ET instrument (Figure 1) makes it possible to
attempt to extract the radial velocity without using simultaneous
calibration with the iodine cell. When the  cell is not inserted
into the optical path, the resulting observations of the star can
explicitly be modelled as the template spectrum with a phase shift
added (the template spectrum may also be the stellar template of a
reference star)
\begin{equation}
V_D e^{i\theta_D} = V_S
e^{i\theta_{S_0}}e^{i\theta_S-i\theta_{S_0}}
\end{equation}
In such a case the phase shift, $\theta_S-\theta_{S_0}$, is a
combination of the velocity shift of the star compared to the
template and  the instrument drift. When using this technique the
ET observing procedure requires all star-only exposure to have an
iodine template taken immediately before or after the exposure. If
the instrument is stable enough, these iodine exposures can be
used to determine and calibrate out the instrument drifts for the
star exposures. This is only possible when the instrumental
velocity drifts are small in the time interval between the
exposure mid-points of the star and iodine templates.

Using the entire spectral range of $\sim640$ \AA\ a
cross-correlation was performed in log-lambda space using the
visibilities ($V$) of the reference star and the target star. The
purpose of this cross-correlation was to yield an approximate
velocity shift, accurate to $\sim 1-2$ km s$^{-1}$, which is
useful as a first guess starting point to calculate a more precise
velocity shift using both the visibility and phase information.

The velocity extraction algorithm that is part of the ET pipeline
was used to minimize the residuals between the complex
visibilities (${\bf V}$) of the reference star and the target star
by varying the phase shift applied to the reference template
complex visibility. The visibility of the reference star was also
allowed to scale in the fitting procedure in an attempt to
minimize the template mismatch, as described in the previous
section. This technique yields a phase shift
$\theta_S-\theta_{S_0}$ that is a measure of the difference in
velocity between the stellar template used and the target star.

To account for instrument drifts between the reference star
template and the target star, a similar analysis procedure was
followed for the iodine templates (taken very close in time to the
reference and target star exposure) with the only difference being
that the iodine visibility was not allowed to scale (since there
is no template mismatch between iodine cell spectra taken at
different times). The phase drift between the two iodine templates
is a direct measure of the instrument drift. Subtracting off the
iodine phase shift corrects for the instrument drift, giving the
real phase shift of the target star with respect to the reference
star.

\subsection{Radial Velocity from Star+Iodine Data}
In the normal data acquisition mode the stellar beam passes
through the iodine cell. Cross-correlation with the visibilities
from a reference stellar template alone is not appropriate since
the iodine spectrum is also present in the data. Using Equation 2
we can write
\begin{equation}
V_D^2 = V_S^2 +V_I^2 +2V_SV_I \cos(\theta_S-\theta_I)
\end{equation}
 The term $2V_SV_I \cos(\theta_S-\theta_I$) has an expectation value close to zero over the
640 \AA\ wavelength band since the cosine term fluctuates
randomly. We ignored this term and approximated the visibilities
as adding in quadrature ($V_D^2 \approx V_S^2 +V_I^2$). This
allowed us to use the TODCOR algorithm (Zucker \& Mazeh 1994) on
the ensemble of measured visibilities to determine the approximate
velocity shifts of the star and iodine templates that are a best
fit to the data. The TODCOR algorithm is a two dimensional
cross-correlation technique generally used to determine the
individual radial velocities of both components of a spectroscopic
binary, and can be applied in this case since the two components
are approximately additive. Since a reference stellar template for
a different star ($\eta$ Cas) was being used, a visibility scaling
was derived using mean visibilities and Equation 6, and used to
scale the stellar visibilities for TODCOR. The scaling factor in
visibilities accounts partially for the mismatch in stellar
templates. The approximate radial velocity values derived using
TODCOR were useful in determining the correct phase shift when the
shift was more than $ \pm 2\pi$. These radial velocities were used
as a first guess to determine the correct wavelength-shift and
phase-shift of the stellar and iodine templates that minimized the
residuals in the complex visibility between the data and the best
fit solution. Subtracting off the iodine phase shift gives the
phase shift of the target star with respect to the reference star,
corrected for instrument drift.

\section{Radial Velocity Results}
The differential radial velocity precision of one of the reference
stars, $\eta$ Cas, is shown in Figure 2. These velocities were
extracted using the techniques described in Section 3, but with an
arbitrary zero-point. In the case of $\eta$ Cas  the velocity
results are identical to those obtained using the standard
pipeline since this object is the reference star. The velocity rms
of $\sim6.4$ m s$^{-1}$ is due to a combination of photon noise
errors in the data and the templates and systematic errors in the
velocity extraction. We used this star as the stellar reference
template to determine barycentric radial velocities. Velocities
for all the stars are calculated relative to $\eta$ Cas, requiring
prior knowledge the barycentric radial velocity of $\eta$ Cas
itself to apply the correct velocity offset. We used the $\eta$
Cas mean barycentric radial velocity from ND02 ($8.314$ km
s$^{-1}$). $\eta$ Cas is known to be a long-period binary, whose
velocity is well fit with a linear trend. We transformed the ND02
velocity to our epoch of observation by applying a correction term
for its known $9.3$ m s$^{-1}$ year$^{-1}$ linear velocity drift
(Cummings et al. 1999). Any error in the determination of this
velocity is a error in the zero-point itself, and will affect all
target stars in a similar way. The phase difference of each star
from the $\eta$ Cas stellar template was converted to velocity
using the phase-to-velocity scaling parameter (${c \lambda}/{2 \pi
d}$). This radial velocity was then corrected to the solar system
barycenter by removing the barycentric velocity contribution of
the telescope, calculated with software written in IDL that use
the JPL ephemeris. These algorithms are accurate to better than
$\sim 2$ m s$^{-1}$. The corrected barycentric radial velocity of
$\eta$ Cas was added to obtain the barycentric radial
velocity of the star.%

\subsection{Radial Velocity Results from Templates Alone}
Barycentric radial velocities were derived using the procedure
outlined in Section 3.1. A comparison of these velocities with
those published by ND02 is shown in Figure 3.
 No attempt has been made to correct the velocities for the known
planetary companions to 51 Peg, 55Cnc and HD49674, which may cause
additional velocity scatter. Our velocities agree well with those
of ND02 and show a scatter of 120 m s$^{-1}$, with a mean offset
of $-79$ $\pm$ 25 m s$^{-1}$. The agreement of our results with
those of ND02 confirms the validity of using the iodine templates
to remove the instrument drift when such drifts are small (Figure
1). For example, the typical instrument drift over the first night
is approximately 10 m s$^{-1}$ over a 15-minute exposure. Such
exposure times with the ET instrument allow the radial velocity of
an 11$^{th}$ magnitude (V=11) star to be determined at this level
of precision (100 m s$^{-1}$).

\subsection{Radial Velocity Results from Star+Iodine Data}
 When the
required precision is high, it is perhaps more appropriate to use
the star+iodine data since the iodine tracks the instrument drift
exactly. For all the reference and planet bearing stars, the
barycentric radial velocities obtained for all data points (using
$\eta$ Cas as template), derived using the procedure outlined in
Section 3.2, were averaged together to give a mean barycentric
radial velocity for each star. These radial velocities are shown
in Figure 4 and display a rms scatter of 47 m s$^{-1}$ with a mean
offset of -3 $\pm$ 19 m s$^{-1}$ in comparison with ND02.

\subsection{Systematic Errors and Velocity Zero Point}
 Tests with twilight solar spectra as the reference stellar template show similar
levels of rms scatter to those using $\eta$ Cas, but the twilight
observation of the solar spectrum (instead of day sky) introduced
an offset in the velocities when the solar spectrum was used as
the template. Our results depend on the velocity of $\eta$ Cas
derived by ND02 who use the National Solar Observatory (NSO) solar
spectrum (Kurucz et al. 1984) as the stellar template with a 522 m
s$^{-1}$ offset to make the radial velocity of the Sun zero. Due
to the use of the same zero point
 our estimated velocities will exhibit similar systematic errors with spectral type as ND02,
 in addition to any template mismatch effects introduced due to our use of
$\eta$ Cas (G0V) instead of the Sun (G2V).

A number of  systematic errors can occur in velocity determination
using this technique. Fast rotators ($v \sin{i} >$ 12 km s$^{-1}$)
have very low fringe visibilities at our chosen interferometer
optical delay (7 mm), leading to a significantly reduced velocity
precision for such stars. Incorrect calibration of the
interferometer delay can lead to errors in the phase-to-velocity
scaling, causing errors in the calculated radial velocity. Our
reference ($\eta$ Cas) has a small linear drift in velocity that
we have accounted for using the mean epoch of observation in ND02.
An error in this determined barycentric radial velocity of $\eta$
Cas will appear as a shift in the mean offsets reported here. The
stars with known planets have additional velocity variability due
to their companions. This variability is reduced when using
star+iodine data (Section 4.2) and averaging all the observed
velocities, especially for the short period planets 51 Peg and
HD49674 that have orbital periods less than our 12 day observing
run. The multi-planet 55 Cnc system (McArthur et al. 2004),
though, has long period planets which affect the calculated radial
velocity. For each star in their sample ND02 report only the mean
of the observed barycentric radial velocities, and to ensure a
consistent comparison we have made no attempt here to correct our
data for these known planetary companions. In general, velocity
corrections for known companions is easy to implement if the
orbital parameters are known, but we have not done so for
consistency with ND02.

\section{DISCUSSION}
We have demonstrated the ability to measure barycentric radial
velocities at a high level of precision (rms = $50 -120$ m
s$^{-1}$) using a dispersed fixed delay interferometer instrument.
Our results also show that the radial velocities derived using a
very different technique and an instrument at substantially lower
resolution are not very different in precision from those derived
using high resolution spectroscopy and cross-correlation with
stellar (ND02; UD99) or Fe I line templates (Gullberg \& Lindegren
2002). Simultaneous calibration with the iodine cell yields more
precise velocities, but the data analysis using only the templates
(Section 3.1, 4.1) may be more appropriate for faint stars where
the low signal to noise may introduce significant errors when
extracting velocities from the star+iodine data. We have corrected
the observed radial velocities to the barycenter of the solar
system. Using the corrected $\eta$ Cas radial velocity from ND02
to set the velocity scale for our radial velocities then
automatically corrects all velocities for the gravitational
redshift and convective blueshift of the Sun (used by ND02 as the
zero point). By adopting the radial velocity of $\eta$ Cas from
ND02 we have automatically adopted ND02 as our velocity reference
for the absolute calibration of a zero-point and performed a
secondary measurement to determine the velocities of all our other
target stars. No attempt is made to further correct the stellar
radial velocities for spectral type and luminosity class dependent
effects. This results in our derived velocities having typical
accuracies of $0.2-0.3$ km s$^{-1}$ (derived from ND02, our
velocity reference) for late F, G and K dwarfs, and precisions of
$\sim 50-100$ m s$^{-1}$ when comparing stars of similar spectral
types. With the ET instrument scheduled to be a facility
instrument at the KPNO 2.1m, this velocity measurement capability
will be available to the wider astronomical community. These
techniques will also be applied to data from the MARVELS survey
that will observe $\sim$ 10,000 stars as part of SDSS III, leading
to precise barycentric radial velocities being measured for the
majority of the stars in the sample.

 Future work with these instruments will focus on using the day sky spectra of the Sun as a stellar
template to determine the velocity zero point, allowing
independent comparison with the radial velocities of ND02 and UD99
without having to adopt either of these works as the velocity
reference, and a better understanding of any zero point
differences that may arise due to the much lower spectral
resolution (Dravins \& Norlund 1990) of ET compared to high
resolution echelle spectrographs.

\acknowledgments We are grateful to Richard Green, Skip Andree,
Daryl Wilmarth and the  KPNO staff for their generous support.
 We would also like to thank the anonymous referee for a careful reading of the manuscript and for very constructive suggestions.
 This work is supported by National Science Foundation grant AST
02-4309, JPL, the Pennsylvania State University, and the
University of Florida. J. v. E. and S.M. acknowledge travel
support from KPNO and the Michelson Fellowship. This research has
made use of the SIMBAD and Vizier databases, operated at ADC,
Strasbourg, France. This work was performed in part under contract
with the Jet Propulsion Laboratory (JPL) funded by NASA through
the Michelson Fellowship Program. JPL is managed for NASA by the
California Institute of Technology. This work is based on data
obtained at the Kitt Peak 2.1m telescope.

\clearpage

\begin{figure}
\includegraphics*[scale=0.8,angle=90]{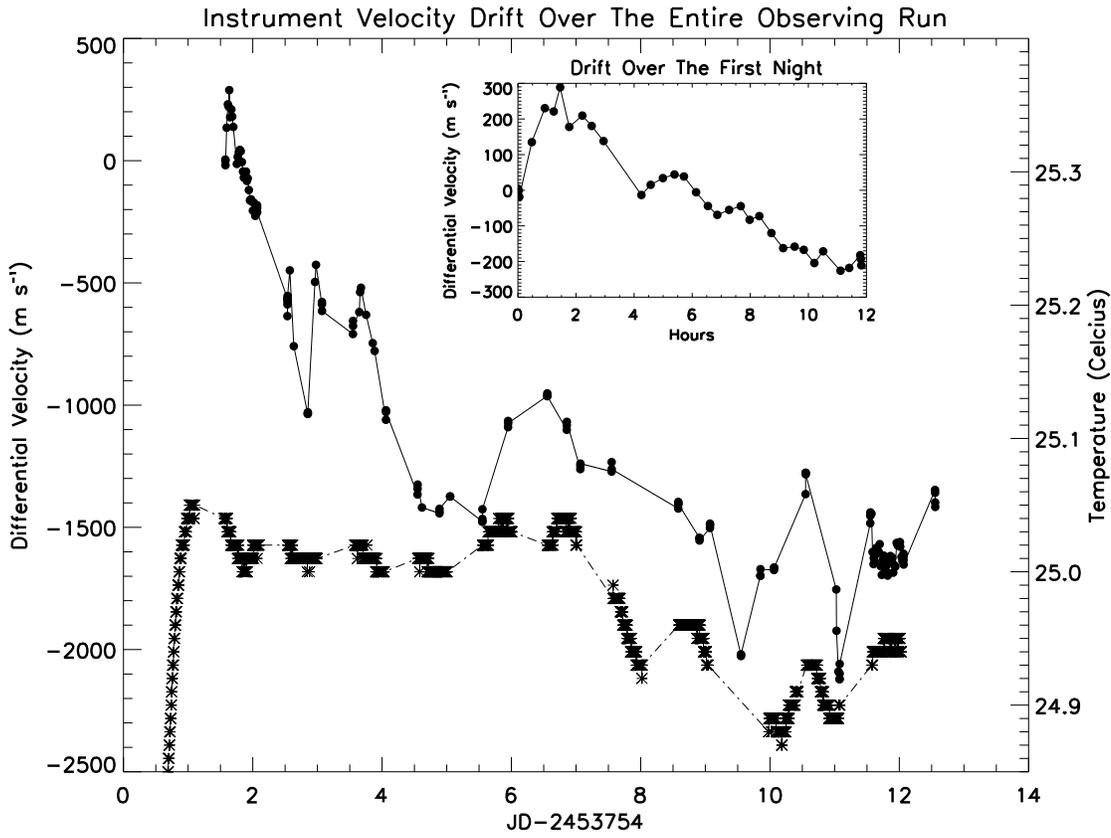}
\caption{Velocity drift of the ET instrument. The solid line is
the velocity drift as tracked by the iodine templates (filled
circles). The dashed line tracks the temperature drift of the
Michelson interferometer (asterisks). A loose correlation between
the temperature drifts and the velocity drifts is evident.
Instrument stability is essential for obtaining radial velocities
if the simultaneous iodine calibration is not used.}
\end{figure}

\begin{figure}
\includegraphics*[scale=0.8,angle=90]{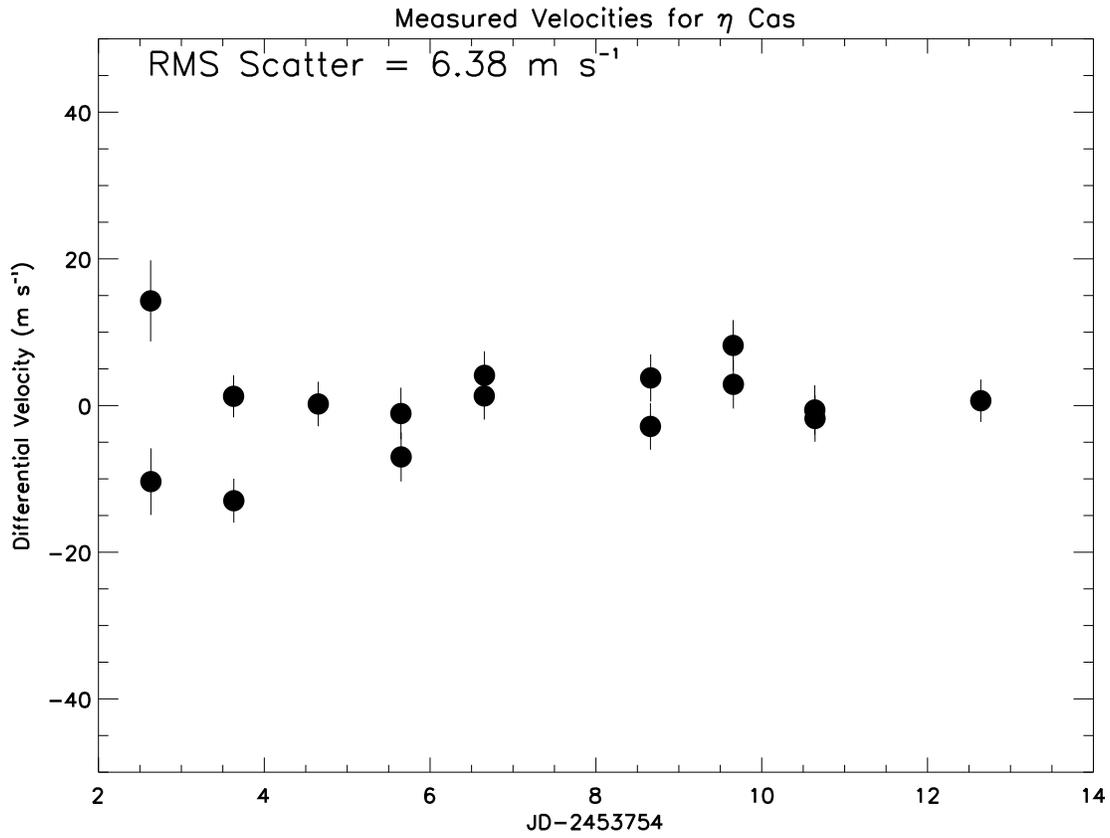}
\caption{Differential radial velocity observations of the star
$\eta$ Cas with ET. The rms velocity scatter of $\sim6.4$ m
s$^{-1}$ is caused by a combination of photon noise errors and
systematic errors. This star is used as the stellar template to
determine barycentric radial velocities (see Section 4).}
\end{figure}

\begin{figure}
\includegraphics*[scale=0.8,angle=90]{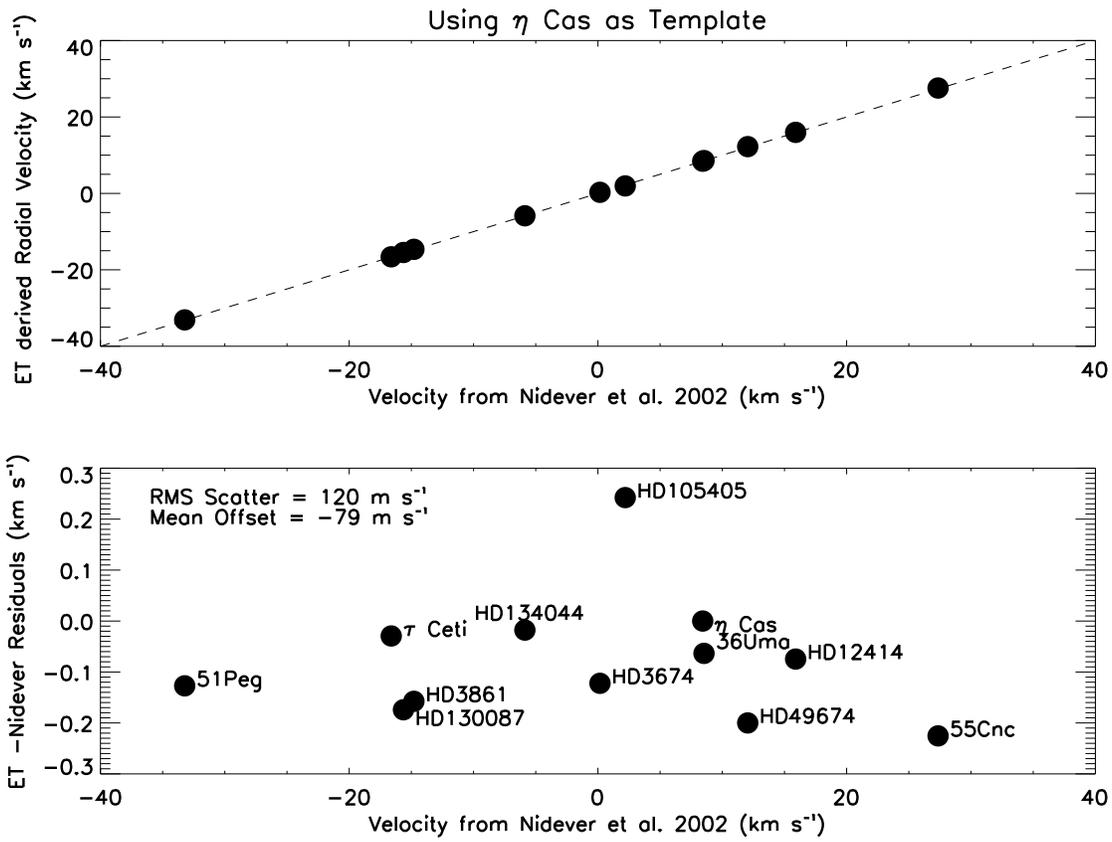}
\caption{Comparison of radial velocities obtained by ET, from
stellar and iodine templates alone, with ND02. The comparison of
ET velocities with ND02 velocities yields an rms velocity scatter
of 120 m s$^{-1}$.}
\end{figure}

\begin{figure}
\includegraphics*[scale=0.8,angle=90]{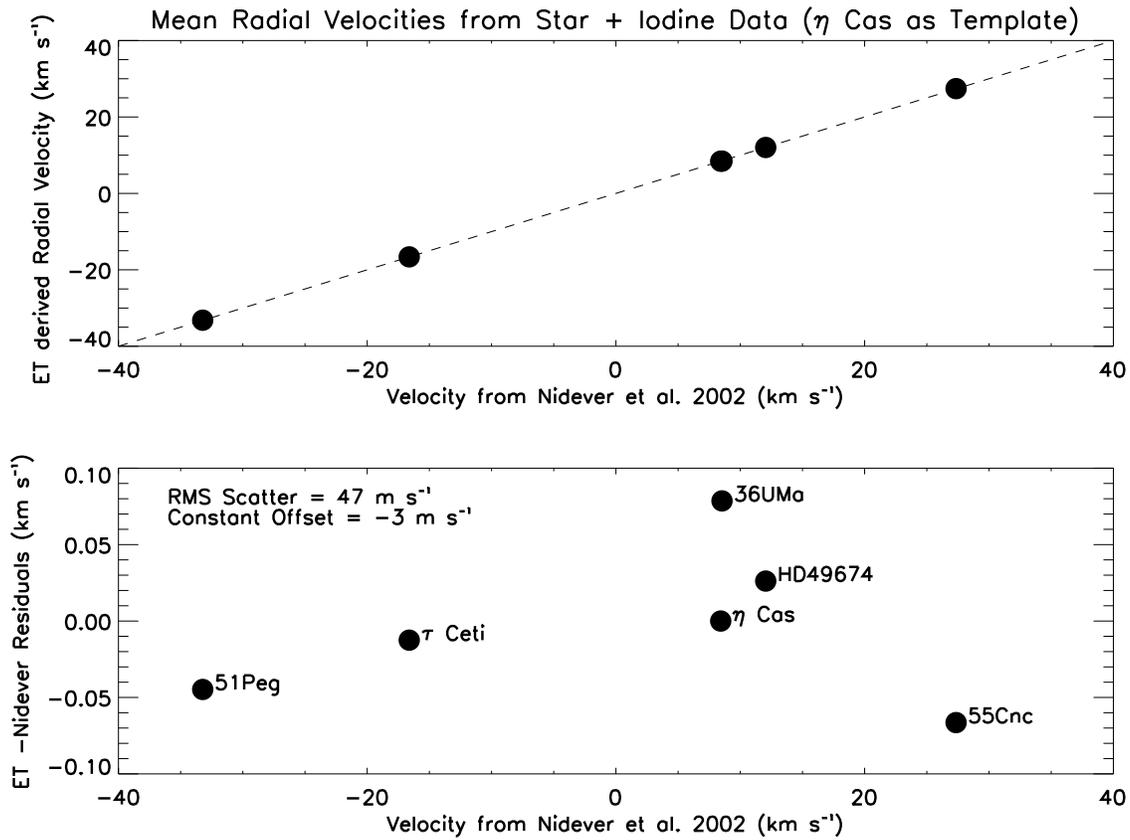}
\caption{Residuals of the average ET radial velocities derived
from star+iodine data from the velocities of ND02. Residuals show
an rms velocity scatter of 47 m s$^{-1}$.}
\end{figure}

\end{document}